\documentclass[slac_one]{revtex4}
\usepackage{graphicx,graphpap}
\usepackage{fancyhdr}
\newcommand\bmat{\left( \begin{array}{cc}}
\newcommand\emat{\end{array}\right)}

\def\msbar{\ifmmode{\overline{\rm MS}} \else{$\overline{\rm MS}$} \fi}
\def\drbar{\ifmmode{\overline{\rm DR}} \else{$\overline{\rm DR}$} \fi}

\def\ti              {\tilde}

\def\b               {\beta}

\def\D               {\Delta}

\def\x               {\chi}

\def\sf              {{\ti f}}

\newcommand{\neu}[1]   {{\ti \x^0_{#1}}}

\begin{document}

\title{{\small{2005 International Linear Collider Workshop - Stanford,
U.S.A.}}\\ 
\vspace{12pt}
Precise predictions for SUSY processes at the ILC} 

%

\author{K.~Kova\v{r}\'{\i}k}
\affiliation{Institut f\"ur Hochenergiephysik der
\"Osterreichischen Akademie der Wissenschaften,\\ A-1050 Vienna,
Austria\\ and Department of Theoretical Physics FMFI UK Comenius
University, SK-84248 Bratislava, Slovakia}
\author{W.~\"{O}ller, C.~Weber}
\affiliation{Institut f\"ur Hochenergiephysik der
\"Osterreichischen Akademie der Wissenschaften,\\ A-1050 Vienna,
Austria}

\begin{abstract}
Recent high precision calculations for production processes of the
SUSY particles at the next linear collider are reviewed. Special
attention is paid to the input parameter definition. Numerical
results for the SPS1a' benchmark point as proposed in the SPA
project, are presented.
\end{abstract}

\maketitle

\thispagestyle{fancy}


\section{Introduction}
The Minimal Supersymmetric Standard Model (MSSM) provides the most
attractive and best studied extension of the Standard Model (SM).
The MSSM predicts the existence of not yet observed supersymmetric
partners to the SM particles. Among those there are the sfermions
$\sf_i$, the scalar partners of the SM fermions, the neutralinos
$\neu{i}$, the fermionic partners of the photon, $Z$-boson and the
neutral Higgs bosons and last but not least the charginos
$\x_i^{\pm}$, the fermionic partners of the $W$-bosons and the
charged Higgs bosons.
\newline %
It is expected that some of these particles will be detected at
the LHC or the Tevatron. To determine their precise properties we
will have to wait for a linear $e^+ e^-$ collider. The future
linear collider should allows measurements with high precision.
After measuring the masses and the couplings of the supersymmetric
particles the main focus will be the extraction of the fundamental
parameters of the MSSM. Although straightforward at tree-level, it
is no longer simple when higher orders are included which are
necessary in view of the precision of the linear collider. The
definition of the parameters is not unique beyond the leading
order and depends on the renormalization scheme. Therefore, a
well-defined theoretical framework has been recently proposed
within the so-called SPA (SUSY Parameter Analysis) project
\cite{SPA}. The "SPA convention" provides a clear base for
calculating masses, mixing angles, decay widths and production
processes. It also provides a clear definition of the fundamental
parameters using the \drbar (dimensional reduction)
renormalization scheme which allows one to extract them from
future data.
\newline %
Here we present the one-loop results for the neutralino, chargino
and sfermion (of the 3rd generation) production processes which
were calculated in the SPA convention mentioned above (for details
and other references see \cite{wcharge,mypaper}). Despite the fact
that the parameters in the SPA convention are defined in the
\drbar renormalization scheme, we have used an on-shell
renormalization scheme. We show how one can obtain results full in
accord with the SPA convention using the on-shell renormalization
scheme. For this purpose, the \drbar parameters are transformed
into on-shell parameters which can then be used as input in the
calculation. This transformation is presented here in some detail.
At the end, we show numerical results for the pair production of
stops, sbottoms, staus, charginos and neutralinos in $e^+ e^-$
collisions.
\section{Parameter transformation}
\begin{figure}
\begin{picture}(178,51)(0,0)
    \put(10,5){\mbox{\resizebox{11cm}{!}
    {\includegraphics{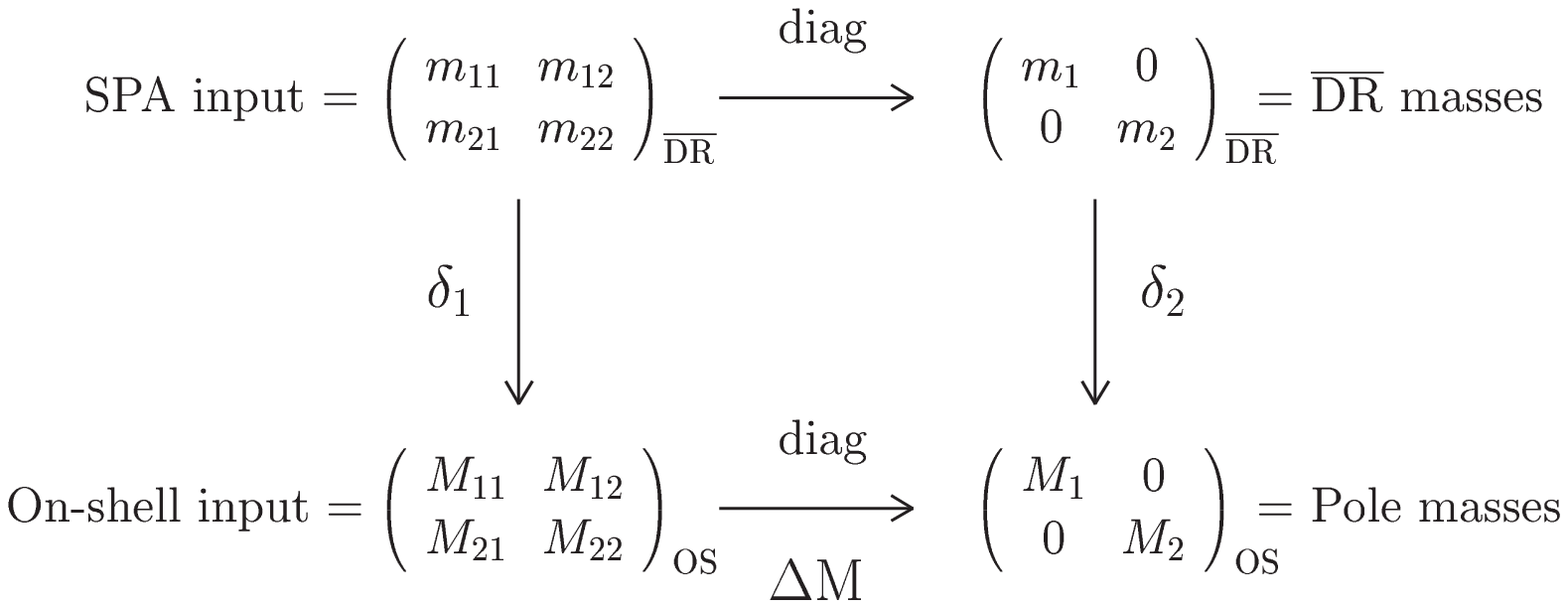}}}}
    \put(135,25){\begin{tabular}{|c||c|c|}
   \hline
  ${\cal P}$ & \drbar & {\rm OS} \\ \hline \hline
  $M'$ & 103.22 & 100.32\\
  $M$ & 193.31 & 197.03\\
  $\mu$ & 402.87 & 399.94\\
  $\tan\b$ & 10 & 10.31\\
  $M_{\tilde{Q}_3}$ & 471.26 & 507.23\\
  $M_{\tilde{U}_3}$ & 384.59 & 410.11\\
  $M_{\tilde{D}_3}$ & 501.35 & 538.92\\
  $M_{\tilde{E}_3}$ & 109.87 & 111.58\\
  $M_{\tilde{L}_3}$ & 179.49 & 181.78\\
  \hline
\end{tabular}}
\end{picture}
\caption{Left: Diagram illustrating the transformation of the
sfermion mass matrix parameters from the \drbar input to the
on-shell input and subsequently diagonalizing the matrix to obtain
the on-shell masses of the sfermions. Right: Table showing the
values of the same parameters in different renormalization schemes
for the SPS1a' benchmark point.}\label{figpara}
\end{figure}

The precision achieved by the linear collider implies the need for
next-to-leading order predictions to be included. In principle one
is free to choose any consistent renormalization scheme for the
next-to-leading order corrections. We show here that by using any
such scheme one can still make a full use of the SPA conventions.
One just has to transform the parameters in a consistent way. From
now on we focus on transforming the parameters from the \drbar to
the on-shell renormalization scheme.
\newline %
As an example, we take the sfermion mass matrix (see
Fig.~\ref{figpara}) which is a non-diagonal matrix at the
beginning. There are two equivalent ways how to obtain the
sfermion pole masses (as one can see in Fig.~\ref{figpara}).
\newline %
First, in the \drbar scheme one diagonalizes the SPA input matrix
to obtain the \drbar masses. To get the pole masses one has to
subtract the finite parts of the on-shell counterterms $\delta_2$.
\newline %
If one uses the on-shell renormalization scheme, one first has to
transform the SPA input matrix into the on-shell input matrix.
This is done by subtracting the finite parts of the on-shell
counterterms which correspond to the parameters found in the mass
matrix. However, there is a subtlety hidden. As some parameters
are common to more than one mass matrix (e.g. $M_{\tilde{Q}_3}$ is
found in the stop and in the sbottom mass matrix), there are more
possibilities how to define these parameters. This results in more
possible counterterms for one parameter which one can use in the
transformation. If one chooses one particular definition for the
parameter (say we define $M_{\tilde{Q}_3}$ in the sbottom mass
matrix) one has to take into account finite shifts $\D M$ for the
same parameter appearing elsewhere (i.~e. we use
$M_{\tilde{Q}_3}+\D M_{\tilde{Q}_3}$ in the stop sector). There
are two different approaches how to define these shifts which can
be found in \cite{oursche,Hollik}, where we follow the approach
given in \cite{oursche}.
\newline %
The advantage of using the on-shell input values is that the
well-established procedure of on-shell renormalization can be
applied.
\newline %
We use the transformation mentioned above for the SPS1a' benchmark
point which we need for the numerical analysis of the production
processes. The table with the values of the parameters in
different schemes can also be found in Fig.~\ref{figpara}.
\section{Production processes}
\begin{figure}
\begin{picture}(178,49)(0,0)
    \put(4,3){\mbox{\resizebox{6.8cm}{!}
    {\includegraphics{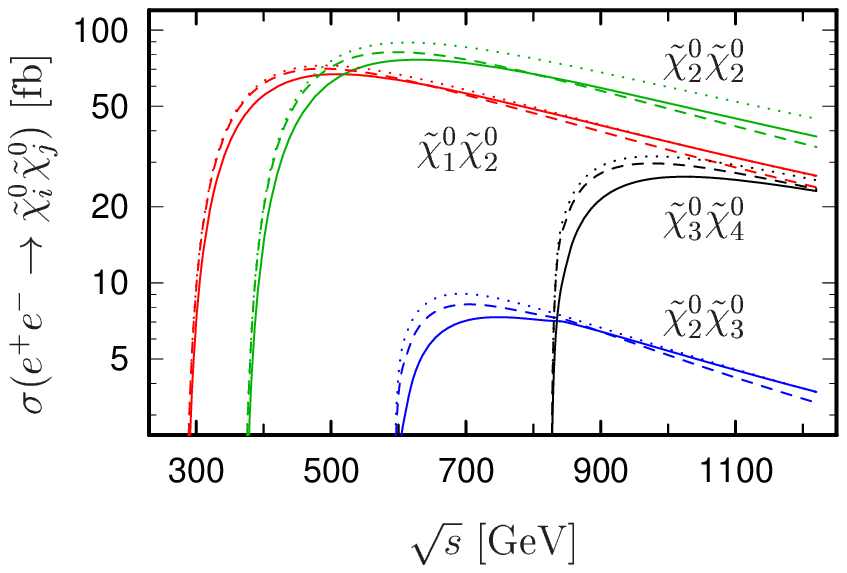}}}}
    \put(90,3){\mbox{\resizebox{6.8cm}{!}
    {\includegraphics{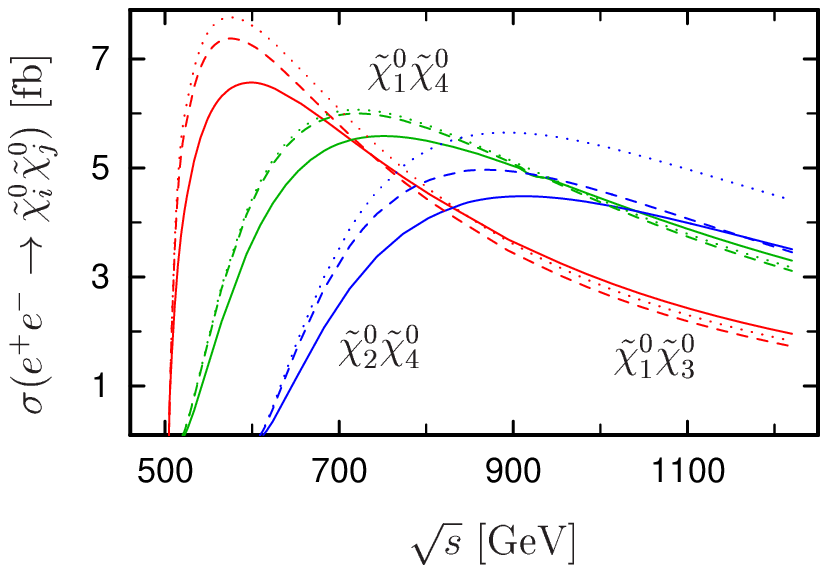}}}}
\end{picture}
\begin{picture}(178,49)(0,0)
    \put(4,2.5){\mbox{\resizebox{6.8cm}{!}
    {\includegraphics{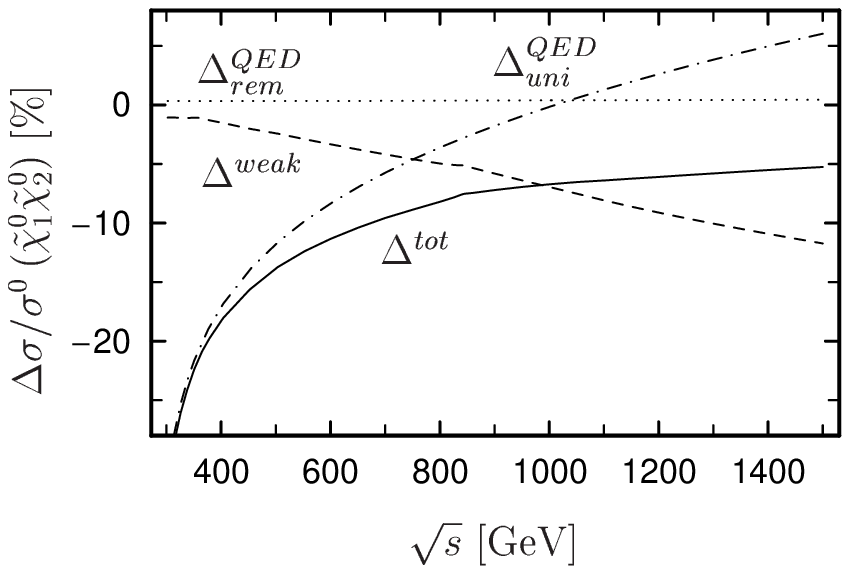}}}}
    \put(90,3){\mbox{\resizebox{6.8cm}{!}
    {\includegraphics{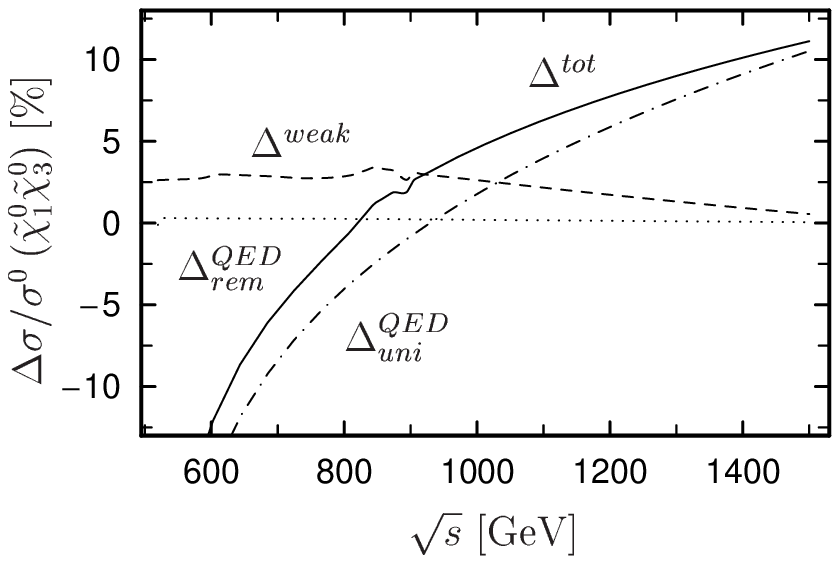}}}}
\end{picture}
\begin{picture}(178,49)(0,0)
    \put(4,-2){\mbox{\resizebox{6.8cm}{!}
    {\includegraphics{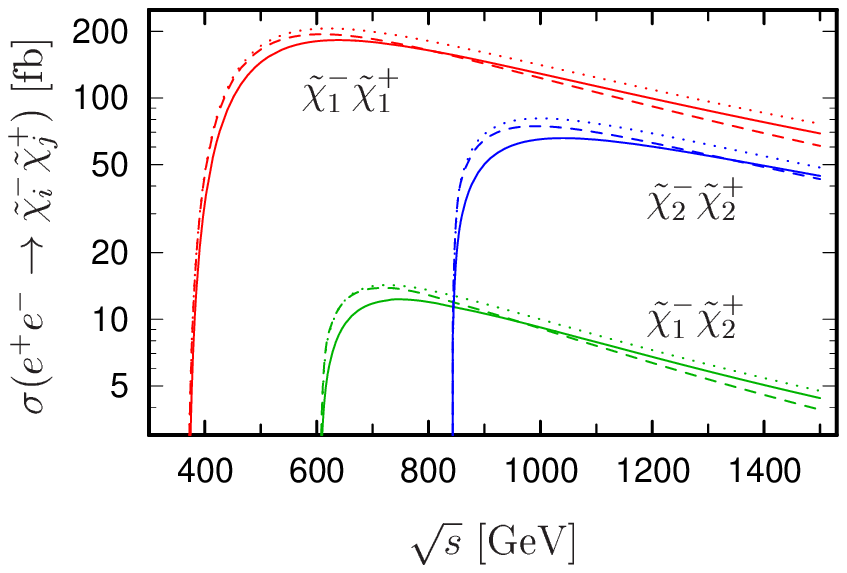}}}}
    \put(90,0){\mbox{\resizebox{6.8cm}{!}
    {\includegraphics{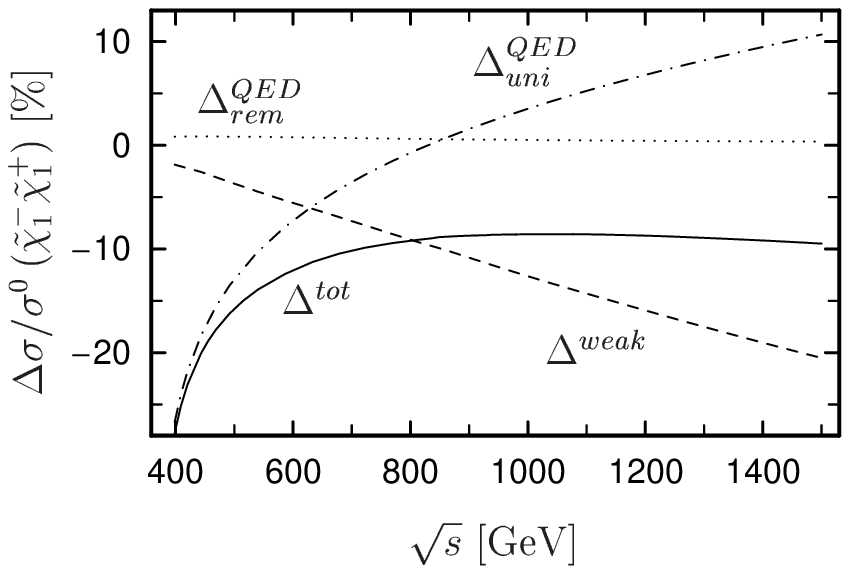}}}}
\end{picture}
\caption{Neutralino and chargino production total cross-sections
for all channels and unpolarized beams accompanied by the relative
corrections for some channels.} \label{figneuchar}
\end{figure}

\begin{figure}[h!]
\begin{picture}(178,54)(0,0)
    \put(11,4){\mbox{\resizebox{8cm}{!}
    {\includegraphics{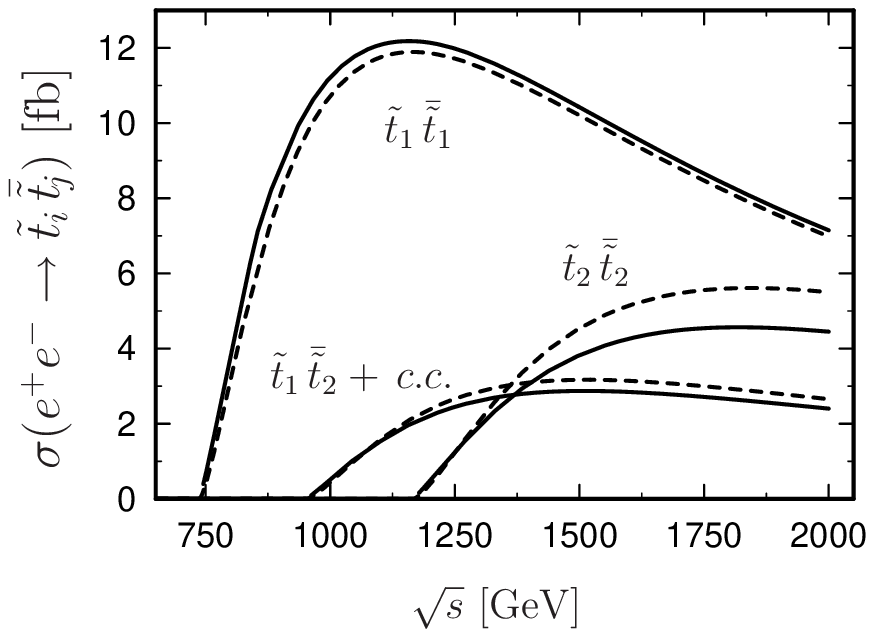}}}}
    \put(97,4){\mbox{\resizebox{8cm}{!}
    {\includegraphics{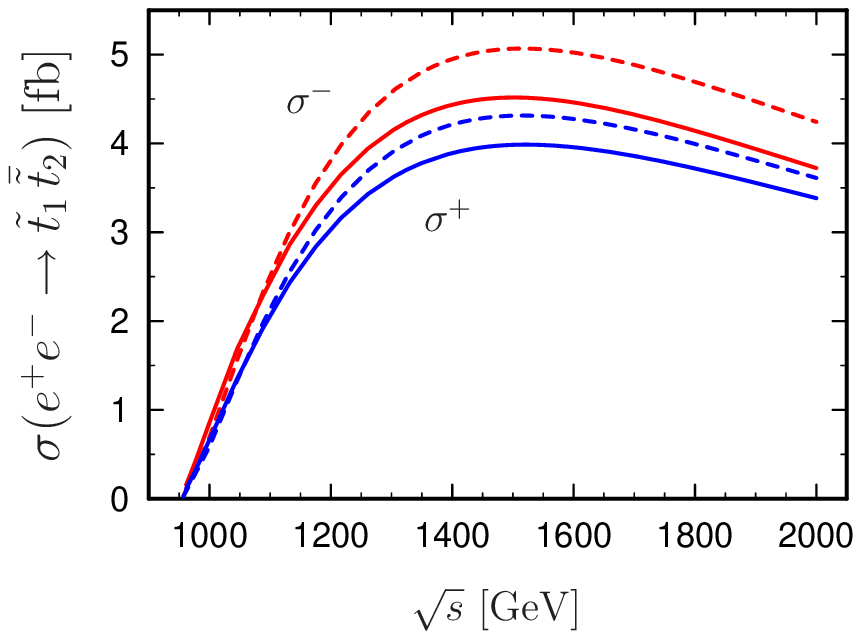}}}}
\end{picture}
\begin{picture}(178,54)(0,0)
    \put(11,2){\mbox{\resizebox{8cm}{!}
    {\includegraphics{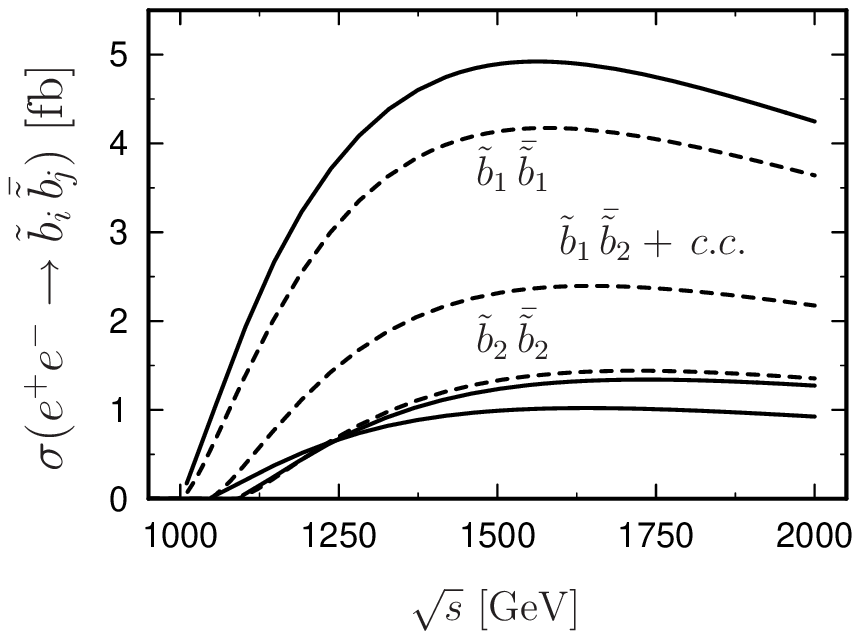}}}}
    \put(97,2){\mbox{\resizebox{8cm}{!}
    {\includegraphics{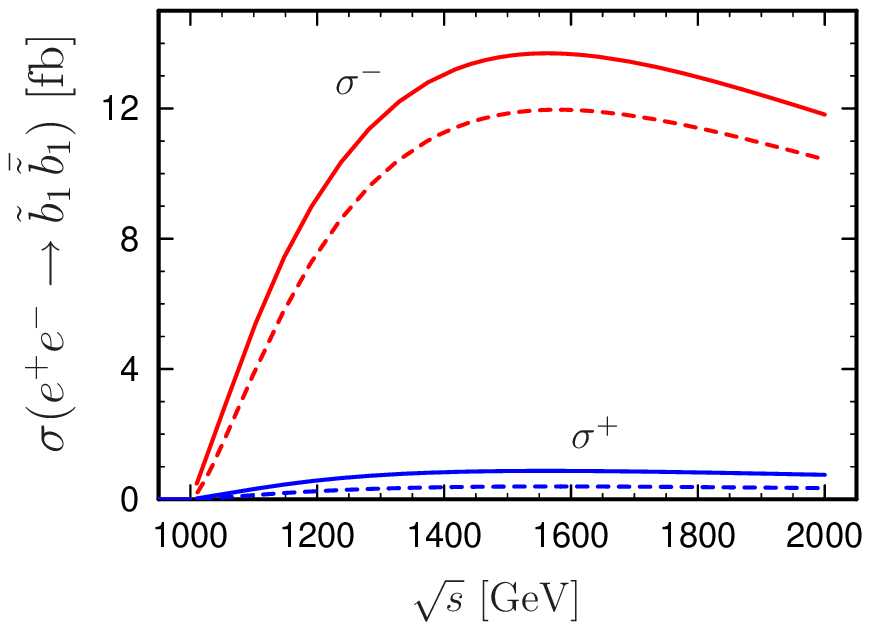}}}}
\end{picture}
\begin{picture}(178,54)(0,0)
    \put(11,0){\mbox{\resizebox{8cm}{!}
    {\includegraphics{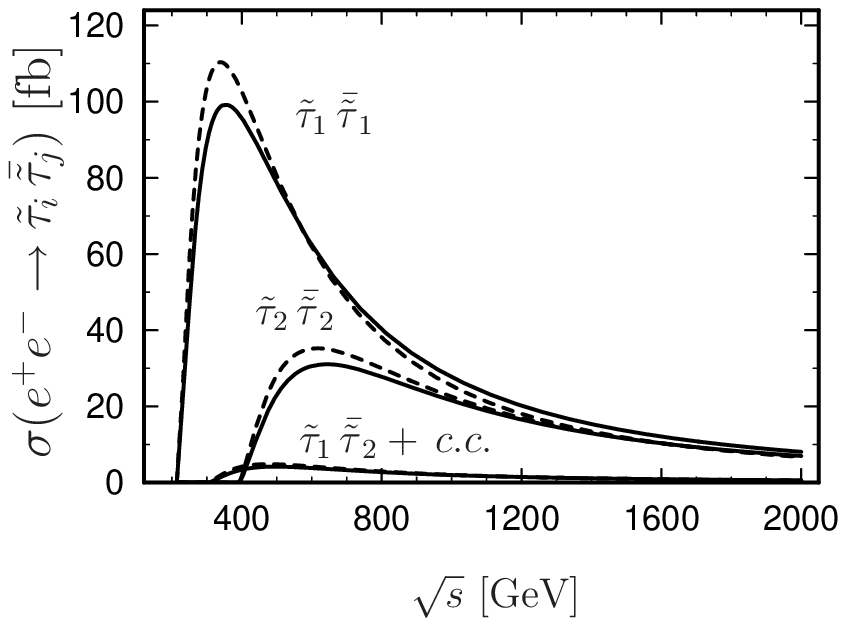}}}}
    \put(97,0){\mbox{\resizebox{8cm}{!}
    {\includegraphics{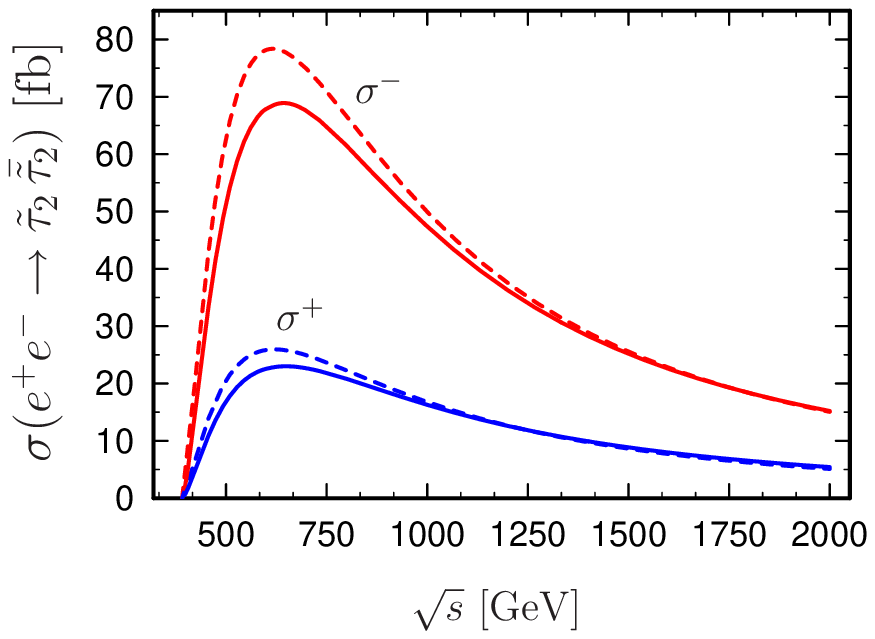}}}}
\end{picture}
\caption{Left: Stop, sbottom and stau production total
cross-sections for all channels and unpolarized beams. Right: The
total cross-sections for polarized beams where $\sigma^-$ stands
for $P_- = -0.8,\, P_+ = 0.6$ and $\sigma^+$ for $P_- = 0.8,\, P_+
= -0.6$ polarizations. The tree-level/complete cross-sections are
denoted by dashed/solid lines.}\label{figsfer}
\end{figure}

The plots in Fig.~\ref{figneuchar} and \ref{figsfer} depict the
total cross-sections of the complete ${\cal O}(\alpha)$ and
leading higher order results for the neutralino, chargino and 3rd
generation sfermion production processes (as presented in detail
in \cite{wcharge,mypaper}). The numerics was done using the SPS1a'
benchmark point and the tree-level results were done also in
accord with the convention i.~e. using \drbar couplings and
on-shell masses. In the case of neutralino and chargino production
extra attention is paid to the problem of the separation of the
QED and weak corrections. The approach applied here is identical
with the one specified in the SPA project (see \cite{SPA}), where
the weak corrections are defined as
\begin{equation}
{\rm d}\sigma^{\rm weak} = {\rm d}\sigma^{\rm virt +
soft}+\frac{\alpha}{\pi}\Big((1-L_e-\Delta_{\gamma})\log\frac{4\Delta
E^2}{s}- \frac{3}{2}L_e \Big)\,{\rm d}\sigma^{\rm tree}
\end{equation}
with $\D_{\gamma}$ taking the cut-off dependent terms from final
state radiation and inital/final state radiation interference into
account (the initial state radiation contribution is contained in
the remaining terms). Using this definition one can separate the
weak and the QED corrections in a gauge invariant and
cut-independent way. The QED corrections are further divided into
two parts, the universal one $\D^{QED}_{uni}$ fully given by a
structure function approach and the remainder $\D^{QED}_{rem}$
(for details see \cite{wcharge}).
\newline %
The sfermion production cross-sections are presented in
Fig.~\ref{figsfer} and the emphasis is put on showing the effect
of the polarization of the $e^+ e^-$ beams. As one can see the
polarization has a significant effect on both the tree-level
cross-sections and the one-loop corrections.

\begin{acknowledgments}
The authors acknowledge support from EU under the
HPRN-CT-2000-00149 network programme and the ``Fonds zur
F\"orderung der wissenschaftlichen Forschung'' of Austria, project
No. P16592-N02.
\end{acknowledgments}

\end{document}